\documentstyle[aps,12pt,epsfig]{revtex}
\begin{document}
\begin{titlepage}
\pagestyle{empty}
\baselineskip=18pt
\rightline{NBI--96--49}
\rightline{September, 1996}
\baselineskip=15pt
\vskip .2in
\begin{center}
{\large{\bf Nucleation of Quark--Gluon Plasma Droplets}}
\end{center}
\vskip .2truecm
\begin{center}
Axel P.~Vischer

{\it Niels Bohr Institute, DK-2100, Copenhagen \O, Denmark.}

\end{center}

\vskip 0.1in
\centerline{ {\bf Abstract} }
\baselineskip=15pt
\vskip 0.5truecm
The energy densities achieved during central collisions of large
nuclei at the AGS may be high enough to allow the formation
of quark--gluon plasma. We propose that most collisions at AGS
energies produce superheated hadronic matter, but in rare events a
droplet of quark--gluon plasma is nucleated. We estimate the
probability of this to occur based on homogeneous and inhomogeneous 
nucleation theory and discuss possible experimental signals.
 
\end{titlepage} 

\baselineskip=15pt
\textheight8.9in\topmargin-0.0in\oddsidemargin-.0in

\section{Introduction}

One of the mysteries of heavy ion physics at Brookhaven National
Laboratory's AGS is: {\it If hadronic cascade event simulators like
RQMD , ARC and ART~[\ref{vischer:cascades}] produce energy densities
approaching 2 GeV/fm$^3$, yet agree with experiment, where is
the quark--gluon plasma?}  After all, numerous estimates of the
onset of quark--gluon plasma agree that it should occur at about
that energy density, and if there is a first order phase transition,
then the onset of the mixed phase would occur at an even lower density.

A possible answer is, that {\it most collisions
at AGS energies produce superheated hadronic matter and are describable
with hadronic cascade simulators, but in rare events a droplet of
quark--gluon plasma is nucleated which converts most of the matter to
plasma.} 

We investigate two possible production mechanisms: 
homogeneous~[\ref{vischer:nucus}]
and inhomogeneous nucleation theory~[\ref{vischer:in}]. 
In the first mechanism we estimate the probability that thermal
fluctuations in a homogeneous superheated hadronic gas could produce
a plasma droplet, and that this droplet was large enough to overcome
its surface free energy to grow.  In the second mechanism we consider another
source of plasma production which is essentially one of
nonthermal origin. We estimate the probability
that a collision occurs between two highly energetic incoming nucleons,
one from the projectile and one from the target,
that this collision would have produced many pions if it had occurred
in vacuum, but because it occurs in the hot and dense medium
its collision products are quark and gluon fields which make a small
droplet of plasma. All our calculations apply to the reaction Au+Au 
at 11.6 GeV/A.

The production of a plasma droplet should have experimental ramifications.
Since the phase transition to the quark--gluon plasma
is occurring so far out of equilibrium we would expect a significant
increase in the entropy of the final state.  This could be seen in the
ratio of pions to baryons or in the ratio of deuterons
to protons.  Along with the increased entropy should come
a slowing down of the radial expansion due to a softening in the
matter, that is, a reduction in pressure for the same energy
density.  Together, these would imply a larger source size and
a longer lifetime as seen by hadron interferometry. Another dramatic
effect is the development of a shoulder in the charged particle
multiplicity distribution. The shoulder becomes more pronounced the
more likely quark--gluon plasma production becomes and therefore the
more charged mesons we produce in this new class of events.

\section{Homogeneous Nucleation Theory}

We need three ingredients to estimate the probability for quark--gluon
plasma production using homogeneous nucleation theory: an equation of
state, the nucleation rate for plasma droplets and a global
description of the dynamical evolution of the nuclear collision.

The dynamics of a central nucleus-nucleus collision at the AGS is
extremely complicated.  We shall be satisfied here with a simple model.
Imagine the colliding nuclei as two Lorentz contracted disks in the
center of momentum frame.  At time $t = 0$ they touch.  They
interpenetrate between $0 \leq t \leq t_0$ where $t_0 = R/v\gamma$,
$R$ and $v$ are the nuclear radius and velocity, and $\gamma$ is the 
Lorentz factor in
the center of momentum frame.  At the end of this time the nuclei
are completely stopped.  The volume of overlap is a linear
function of time.

After the time $t_0$ the hot fireball expands radially.  At late
times we would expect its radius to grow linearly with time.
Therefore we parametrize the volume as $V(t)~\sim~(t~+~c)^3$.
The constant $c$ and the proportionality factors are determined by matching 
the two functional forms for the volume and their first derivatives
at $t_0$. 

To determine the equation of state for baryon rich matter we assume a first
order phase transition. The quark-gluon plasma phase is described by a free 
gas of massless quarks~(u,~d,~s) and gluons confined by a bag constant 
B = (220 MeV)$^4$, while the hadronic phase
consists of a gas of mesons ($\pi$, $K$, $K^{*}$, $\eta$, $\eta^\prime$, 
$\rho$, $\omega$, $\phi$ and $a_1$) and baryons ($N$, $\Delta$, $\Lambda$ and
$\Sigma$) interacting via a repulsive
mean field of strength K = 1500 MeVfm$^3$.
Both phases are joined through a Maxwell construction. 

Given the lab kinetic energy of the collision we can determine the initial
energy density and baryon density in the overlap region and
the starting point of the superheated hadronic system in the $T-\mu_B$ plane.
Assuming baryon number conservation and an entropy conserving hydrodynamic 
expansion we determine 
the time evolution of the system in the $T-\mu_B$ plane.  The system
reaches the coexistence curve after a time $t_f$ = 7 fm.
                                               
The rate $I$ to nucleate droplets of quark--gluon plasma in a hadronic
gas per unit volume per unit time is given by 
\begin{equation}
I= I(\mu_B, T)= I_0 \; \exp(-\Delta F_*/T) \, .
\label{eqt1}
\end{equation}
Here $I_0$ is the prefactor and $\Delta F_*$ is the change in free energy
of the system due to the formation of a single critical size droplet of
plasma.

The nucleation process is driven by statistical fluctuations which
produce droplets of quark--gluon plasma in the hadronic phase.
The size of these fluctuations is determined by the free energy difference of
the hadronic phase with and without the plasma droplet. 
The system under discussion is in a superheated state so that the
interplay between the pressure difference and the surface free energy
results in a maximum of the free energy difference at the critical
radius $R_*(T,\mu_B)$.
Droplets with a radius larger than $R_*$ will expand into the hadronic phase,
while droplets with a radius smaller than $R_*$ will collapse.  $\Delta F_*$
is the activation energy needed to create a droplet of critical size $R_*$.

The prefactor $I_0$ is linearly proportional to the dissipative
coefficients, as expected for linear viscous hydrodynamics.  
For the droplets to grow beyond the critical
radius, latent heat must be carried to the surface of the droplet. This
is achieved through thermal conduction and/or viscous damping.
             
Nucleation begins when the two nuclei first collide with each other
and a superheated overlap region is created. It ends when the
expanding system reaches the phase coexistence curve at time $t_f$. 
The average droplet density can be computed from the expression
\begin{equation}
n_{drop}(t) =  \int^{t}_{0} dt' I(\mu_B(t'),T(t'))  \, .
\label{vischer:n}
\end{equation}
The maximum possible value reached by the
droplet density at $t_f$ is $n_{\rm drop} = 2 \times 10^{-5}$
fm$^{-3}$. The volume of
the expanding system, on the other hand, is on the order of
$2 \times 10^3$ fm$^3$.  The average number of
droplets nucleated is thus rather small, roughly 1/100 for a central Au + Au
collision at the AGS.                         
We found furthermore, that the volume fraction $q$ of quark-gluon
plasma is less or on the order of $10^{-2}$.
Finally, we calculated the average radius of a droplet produced $\bar{R}$ and
found the average radius to be less or on the order of
5.0 fm at the time $t_f$ when the system hits the coexistence curve.

We see that in an average central collision at AGS we produce hardly any
quark-gluon plasma ( $q$ is very small) and hardly any plasma droplets are
created ( $n_{\rm drop}$ is very small). But, if a droplet should be
produced it
will grow and fill a sizable part of the system ($\bar{R}$ is large). 
We conclude that perhaps in one out of every 100 or 1000 central
collisions a sizable fraction of the system has undergone the phase 
transition into quark-gluon plasma.

\section{Inhomogeneous Nucleation Theory}

In inhomogeneous nucleation theory we study the initial stage of the
heavy ion collision. A small but growing overlap region
consisting of stopped, hot and dense hadronic matter has already
formed, but additional
target and projectile matter is still colliding with this hot and
dense zone. We are specifically interested in the
possibility that an incoming projectile
nucleon suffers little or no energy loss during its passage
through the hot and dense zone
where it encounters a target nucleon which also has suffered
little or no energy loss.  The energy available in the ensuing
nucleon-nucleon collision, $\sqrt{s}$, can go into meson production.
Suppose that a large number of pions would be produced if the collision had
happened in free space.  Clearly, the outgoing quark and gluon fields
cannot be represented as asymptotic pion and nucleon states immediately.
The fields must expand and become dilute enough to be called real
hadrons.  If this collision occurs in a high energy density medium,
the outgoing quark-gluon fields will encounter other hadrons before
they can hadronize.  It is reasonable to suppose that this ``star
burst" will actually be a seed for quark-gluon plasma formation
if the surrounding matter is superheated hadronic matter.  

A fundamental result from kinetic theory is that the number of
scattering processes of the type 1 + 2 $\rightarrow X$ is given by
\begin{equation}
N_{1+2 \rightarrow X}
= \int dt \int d^3 x \int \frac{d^3 p_1}{(2 \pi )^3} \, f_1 ({\bf x},
{\bf p}_1,t) \int \frac{d^3 p_2}{(2 \pi )^3} \, f_2 ({\bf x},{\bf p}_2,t)
\, v_{12} \, \sigma_{1+2 \rightarrow X}(s_{12}) \, .
\label{vischer:kinetic}
\end{equation}
Here $v_{12}$ is a relative velocity and the $f_i$ are phase space
densities.  A differential distribution in the variable $Y$ is
obtained by replacing the interaction crossection $\sigma$ with $d\sigma/dY$.

For our purpose it is reasonable to represent the colliding
nuclei as cylinders with radius $R$ and thickness $L$.  All the
action is along the beam axis.  We assume that the phase space
distributions are independent of transverse coordinates $x$ and
$y$ and of transverse momentum. The phase space density of nucleon $i$
can then be rewritten as
\begin{equation}
\frac{dp_{zi}}{2\,\pi} \, f_i (z,p_{iz},t) =
\gamma \, n_0 \, \frac{dx_i}{x_i} \sum_{N_i = 0}^{\infty}
\, H(x_i,N_i) \, S(N_i,d_i(z,t)) \, ,
\end{equation}
where $n_0$ is the average baryon density in a nucleus, about 0.145
nucleons/fm$^3$.  Here,
$H(x,N)$ is the probability that the nucleon has momentum fraction
$x$ after making N collisions and
$S(N,d)$ is probability that the nucleon has made $N$ collisions
after penetrating to a depth $d$.
$\sum_{N=0}^{\infty} H(x,N) S(N,d)$ is then the probability that the nucleon
has momentum fraction $x$ after penetrating to a depth $d$.

For the survival function $S(N,d)$ we assume that the collisions
suffered by the nucleons are independent
and can be described by a Poisson distribution characterized by the mean free
path $\lambda \sim 0.4 fm$ of nucleons in the hot and dense hadronic matter.
The invariant distribution function $H(x,N)$ describes the momentum degradation
of a nucleon propagating through the hot zone.  Csernai and 
Kapusta~[\ref{vischer:ck}]
evaluated $H$ under the assumption that the nucleons experience
deceleration by sequential scattering. Their result depends on one
parameter $w$, the probability that the nucleon collides
inelastically.  Using $w \sim 0.5$ allowed them to obtain a good
representation of p+A and n+A data with
beam energies in the range of 6-405 GeV. 

The spectrum $dN/ds$ is plotted in the Figure to the left.  
The nucleon-nucleon 
collisions may be referred to as primary-primary,
primary-secondary, and secondary-secondary, depending on whether
the nucleons have scattered from thermalized particles in the
hot zone (secondary) or not (primary).
  
\begin{figure}
\begin{center}
\mbox{\epsfig{file=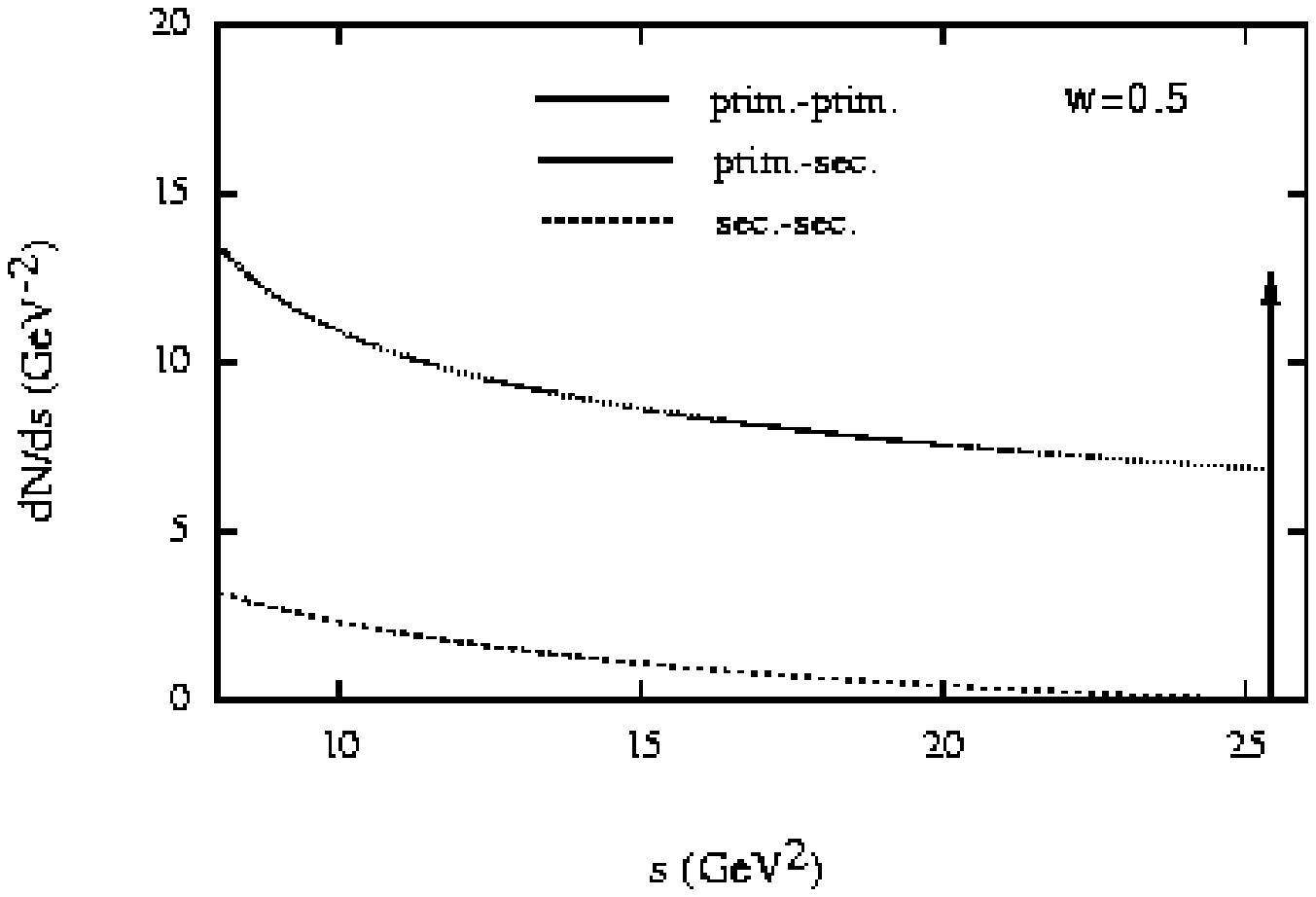,width=8.5cm,height=5.8cm}
\epsfig{file=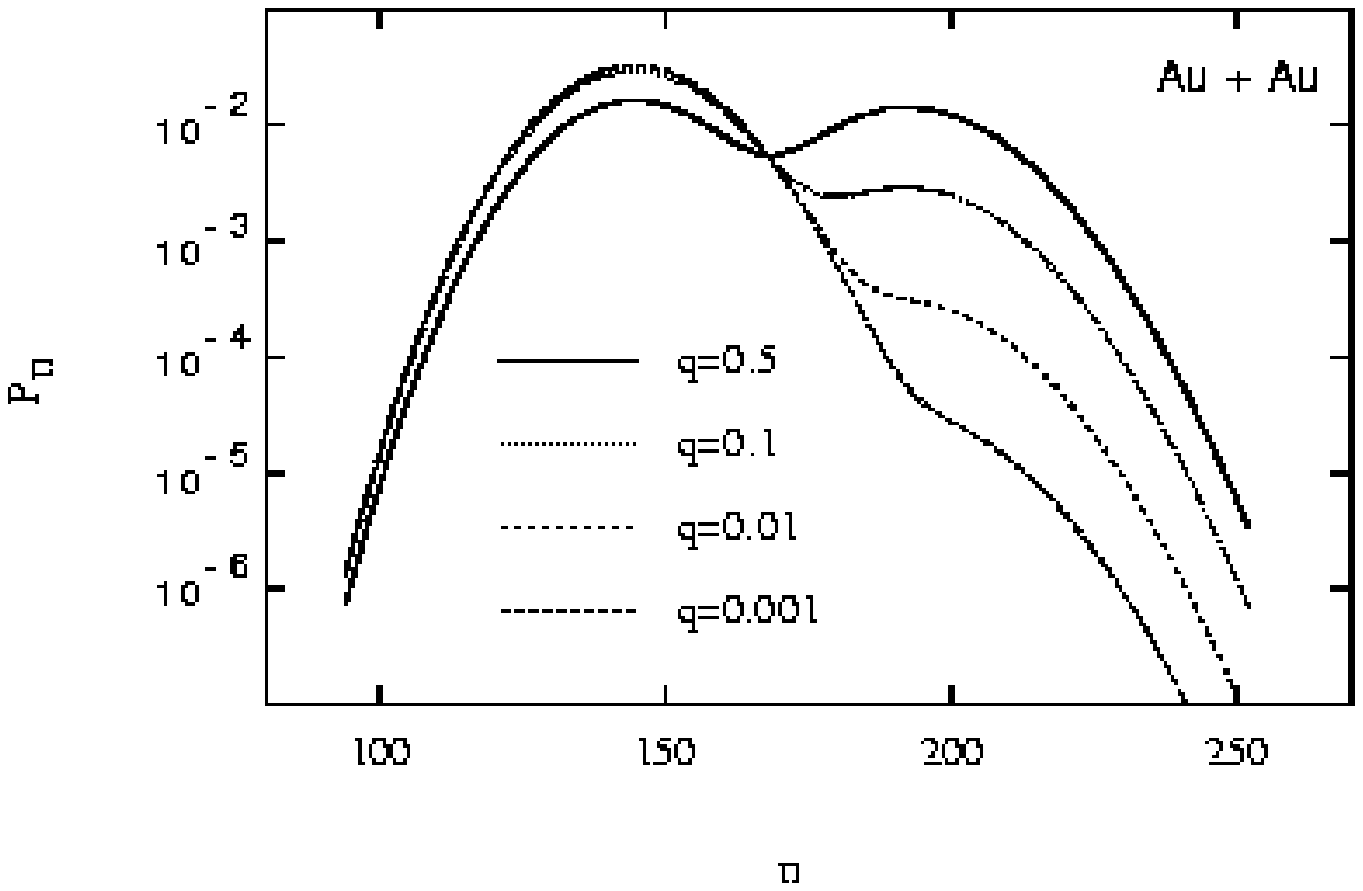,width=8.5cm,height=5.8cm}}
\end{center}
\caption{Left: Distribution $dN/ds$ of hard nucleon--nucleon 
collisions taking place in the hot zone. Right: Negatively charged
multiplicity distribution for different values of $q$}
\end{figure}
Until now we have only selected nucleon-nucleon scatterings
in which the total available energy $\sqrt{s}$ is large.
In addition, we need to specify what fraction of this energy
goes into meson production.
This is described by the pion number distribution function
$P_n (s)$, which is the probability of producing $n$ pions in a
nucleon--nucleon collision in free space. 
Given $P_n (s)$ we can estimate the number of nucleon-nucleon collisions
that would lead to the production of $n$ pions as
\begin{equation}
N_n = \int_{s_{\rm min}}^{4E_0^2} ds\,P_n (s) \,
\frac{dN}{ds}(s)\,.
\label{vischer:number}
\end{equation}
The lower limit of integration is fixed by kinematics and
the upper limit by the beam energy.

We shall approximate the pion number distribution function
$P_n(s)$ with a binomial.
\begin{equation}
P_n (s) = \left( \begin{array}{c} n_{\rm max}\\n \end{array} \right)
\xi^n
(1-\xi)^{n_{\rm max}-n}
\label{vischer:bino}
\end{equation}
The parameter $\xi$ is related to the mean multiplicity by
\begin{equation}
\xi(s) = \frac{\langle n(\sqrt{s}) \rangle}{n_{\rm max}(\sqrt{s})}
\end{equation}
Here $\langle n \rangle$ is the average pion multiplicity averaged over
$pp$, $pn$ and $nn$ collisions and can be taken from experimental
data. The maximum number of pions produced in a nucleon-nucleon collision
$n_{\rm max}$ is determined by kinematics. $N_n$ allows us to
evaluate the number of hard nucleon--nucleon collisions, producing $n$
pions in free space. This quantity is our measure for seeds or
nucleation sites of quark--gluon plasma production.

We are interested in the possibility that one of these seeds
nucleates plasma.  The precise criterion for this to
happen is not known.  However, we can make some reasonable estimates.
In [\ref{vischer:nucus}] we estimated that a critical size plasma droplet at
these temperatures and baryon densities would have a mass of about
4 GeV.  Any local fluctuation more massive than this would grow
rapidly, converting the surrounding superheated hadronic matter to
quark-gluon plasma.  Another estimate is obtained
by the argument that at these relatively modest beam energies most
meson production occurs through the formation and decay of baryon
resonances.  The most massive observed resonances
are in the range of 2 to 2.5 GeV.  Putting two of these in close
physical proximity leads to a mass of 4 to 5 GeV.
Let us assume next that each particle, nucleon and meson, carries away
a kinetic energy equal to one half its rest mass.  If a particle
would have too great a kinetic energy then it might escape from
the nucleon-nucleon collision volume long before its neighbors and
so would not be counted in the rest mass of the local fluctuation.
Taking 4 GeV, dividing by 1.5, and subtracting twice the nucleon mass
leaves about 7 pion rest masses.  So our most optimistic estimate
is that one needs a nucleon-nucleon collision which would have led
to 7 pions if it had occurred in free space.  One might be less
optimistic and require the production of 8 or 10 pions instead.

Conservatively, we concluded by evaluating $N_n$ for $n \geq 7$, 
that the probability of at least
one plasma seed appearing via this mechanism is in the range of
1 to 10\% per central gold-gold collision at the highest energy
attainable at the AGS. 

\section{Experimental Signals}

Both homogeneous and inhomogeneous nucleation theory predict, that in
rare events quark--gluon plasma can be formed. In both
scenarios plasma droplets are formed, but with a rather large uncertainty
in both, the parameters and assumptions that go into the theory as
well as in the probabilities we obtained. The final answer about the
likelihood of rare events of quark--gluon plasma production has to
come from experiments.

If a phase transition to quark--gluon plasma occurs in the collision,
we can expect a significant additional production of entropy.  As a
result we should find a strong increase in the meson production. In
general we would expect two distinct two classes of events. One purely
hadronic and the rare events, in which a quark--gluon plasma droplet
was formed.

These two distinct classes of events might be visible in the charged
particle multiplicity distribution which would have the form
\begin{eqnarray}
P_n = (1-q) \, P_{\; n}^{\rm had} (N_{\rm had})
	+ q \, P^{\rm qg}_{\; n} (N_{\rm qg})\, .
\label{vischer:double}
\end{eqnarray}
Here $q$ is the probability of finding a central event in which plasma
is formed, $P_{\; n}^{\rm had}$ is the multiplicity distribution
for purely hadronic events with mean $N_{\rm had}$, and $P^{\rm qg}_{\; n}$
is the multiplicity distribution for events in which a plasma
was formed with mean $N_{\rm qg}$.

To obtain a feeling for the shape and applicability of equation 
(\ref{vischer:double}) we plot in the Figure to the right different negatively
charged particle multiplicity distributions. 
The mean for purely hadronic events is estimated to be
$N_{\rm had} = 145$ and an upper limit of $N_{\rm qg} = 193$ was found
for the rare events, under the assumption that all of the matter is converted
into plasma [\ref{vischer:in}]. We use Poisson
distributions for $P^{\rm had}$ and $P^{\rm qg}$ and plot the
negatively charged particle multiplicity distribution defined in eq.
(\ref{vischer:double}) for different values of the probability $q$.
A shoulder develops for small $q$ and becomes more pronounced the
larger $q$ is.

In summary we propose that the formation of plasma in rare events
should have observable consequences for hadron interferometry,
deuteron production, and the meson multiplicity distribution.
For the multiplicity distribution one would observe a shoulder or second
maximum at some multiplicity higher than the most probable one.
If there is a phase transition but it is second order or weakly
first order then the effect will be much more difficult to see.
We eagerly await the results of experiments.

\section{References}

\begin{enumerate}

\item \label{vischer:cascades} 
	H. Sorge (for RQMD), S. Kahana (for ARC) and B.--A. Li (for
	ART), these proceedings

\item \label{vischer:nuc}
	J.S. Langer, Ann. Phys. (N.Y.) {\bf 54}, 258 (1969);
	L.P. Csernai and J.I. Kapusta, Phys. Rev. C {\bf 46}, 1379
	(1992)

\item \label{vischer:nucus}
	J.I. Kapusta, A.P. Vischer, R. Venugopalan, Phys. Rev. C {\bf
	51}, 901 (1995); R. Venugopalan and A.P. Vischer, Phys. Rev. E
	{\bf 49}, 5849 (1994); A.P. Vischer, Nuc. Phys. A {\bf 590}
	585c (1995)

\item \label{vischer:in}
	J.I. Kapusta and A.P. Vischer, Phys. Rev. C {\bf 52} (1995)

\item \label{vischer:ck} 
	L. P. Csernai and J. I. Kapusta Phys. Rev. D {\bf 29},
	2664 (1984); ibid. {\bf 31}, 2795 (1985).
\end{enumerate}

\end{document}